\documentclass{aa}  

\usepackage{graphicx}
\usepackage{txfonts}

\usepackage{pdflscape}	
\usepackage[utf8]{inputenc}
\usepackage[english]{babel}

\usepackage[dvipsnames]{xcolor}
 
\usepackage{graphics}
\usepackage{graphicx}

\usepackage{threeparttable}
\newcommand{\ka}{K${\alpha}$~}

\newcommand{\ltsima}{$\buildrel < \over \sim$}
\newcommand{\lsim}{\lower.5ex\hbox{\ltsima}}
\newcommand{\gtsima}{$\buildrel > \over \sim$}
\newcommand{\gsim}{\lower.5ex\hbox{\gtsima}}

\newcommand{\swift}{{\it Swift}\xspace}

\newcommand{\nus}{{\it NuSTAR}\xspace}

\newcommand{\src}{Swift J1845.7-0037\xspace}
\newcommand{\fdcut}{\texttt{FDCUT}\xspace}

\begin{document}

   \title{ {\it NuSTAR} X-ray spectrum of Be-X-ray pulsar Swift J1845.7-0037 }
   \subtitle{Bulk and thermal Comptonization of cyclotron seed photons in the accretion column}

   \author{Filippos Koliopanos\fnmsep\thanks{Corresponding author, fkoliopanos@irap.omp.eu}
          \inst{1,2}
          \and
          {Georgios Vasilopoulos}\inst{3}
          \and{Michael T.~Wolff}\inst{4}  
          }

   \institute{CNRS, IRAP, 9 Av. colonel Roche, BP 44346, F-31028 Toulouse cedex 4, France\\\
              \email{fkoliopanos@irap.omp.eu}
         \and
             Universit{\'e} de Toulouse; UPS-OMP; IRAP, Toulouse, France 
              \and
              Department of Astronomy, Yale University, PO Box 208101, New Haven, CT 06520-8101, USA 
              \and
              Space Science Division, U.S. Naval Research Laboratory, Washington, DC 20375, USA}

   \date{Received September 15, 1996; accepted March 16, 1997}

 
  \abstract
   {}
   {Spectral and temporal analysis of the \nus observation Galactic Be-XRB \src, during its recent outburst.}
   {For the spectral analysis we use both phenomenological and physics-based models. We employ an often used empirical model to identify the main characteristics of the spectral shape in relation to nominal spectral characteristics of X-ray pulsars. Additionally, we used the latest version of Bulk \& Thermal comptonization model (BW), to assess the validity of the spectral components required by the empirical model and to investigate the origin of the hard X-ray emission. We also analyzed the source light-curve, studying the pulse shape at different energy ranges and tracking the spectral evolution with pulse phase by using the model independent hardness ratio (HR). }
   {We find that while both the empirical and physical (BW) spectral models can produce good spectral fits, the  BW model returns physically plausible best-fit values for the source parameters and does not require any additional spectral components to the non-thermal, accretion column emission. The BW model also yielded an estimation of the neutron star magnetic field placing it in the $10^{12}$\,G range. }
   {Our results, show that the spectral and temporal characteristics of the source emission are consistent with the scattering processes expected for radiation dominated shocks within the accretion column of highly magnetized accreting neutron stars. We further indicate that physically-derived spectral models such as BW, can be used to tentatively infer fundamental  source parameters, in the absence of more direct observational signatures.}


   \maketitle
%

\section{Introduction}

Powered by mass accretion onto highly magnetized ($B{>}10^{10}$\,G) neutron stars (NSs), X-ray pulsars (XRPs) are some of the most luminous (off-nuclear) X-ray point-sources \citep[e.g.,][]{2017Sci...355..817I} in the universe. These are binary star  systems in which material lost by a companion star -- ranging from low-mass white dwarfs to massive B-type stars -- is accreted onto the high-B NS and interacts with its immense magnetosphere \citep[e.g.,][and references therein]{2012MmSAI..83..230C,2015A&ARv..23....2W,2016arXiv160806530W}. Ionized gas "trapped" by the magnetic field lines is channeled towards the poles of the neutron stars, forming a funnel-like structure known as the accretion column.  Material inside the accretion column is heated to high energies \citep[e.g.,][]{1975A&A....42..311B,1976MNRAS.175..395B,1985ApJ...299..138M}, producing copious X-ray photons, which due to intense and highly anisotropic scattering are collimated into a beam parallel to the magnetic field axis \citep[scattering cross section is significantly reduced in the direction of the strong B-field axis][]{1971PhRvD...3.2303C,1974ApJ...190..141L}. A tilt between the rotational axis of the pulsar and the pulsar beam, produces the characteristic pulsations of the X-ray light curve, which accurately track the NS spin. 

Perhaps more remarkably, the combination of beaming of the X-ray emission and guiding of the accretion flow -- both of which are promoted by the strong magnetic field -- can allow and sustain accretion above the (local) Eddington limit \citep{1976MNRAS.175..395B,1981ApJ...251..288N}. The significance of super-Eddington accretion onto high-B NSs, has been further bolstered with the recent discoveries of pulsating ultraluminous X-ray sources (ULXs)\citep[e.g.,][]{2014Natur.514..202B,2016ApJ...831L..14F,2017Sci...355..817I}. ULXs are persistent X-ray point sources with an apparent luminosity well above the Eddington luminosity of stellar mass black holes (i.e.~${L_{X}{\gtrsim}}$a\,few\,$10^{39}$\,erg/s. Initially though to be hosts of the elusive IMBHs, it is gradually acknowledged  that a large fraction -- if not most-- of known ULXs are potentially powered by high-B NSs rather than black holes\citep{2016MNRAS.458L..10K,2017MNRAS.468L..59K,2017MNRAS.467.1202M,2017A&A...608A..47K,2018ApJ...856..128W}. This emerging new paradigm, was motivated by the discovery of the pulsating ULXs (PULX), which confirmed the presence of a highely magnetized NS in these source, and was further promoted by the realization the both pulsating and non pulsating ULXs have disitinct spectral similarities, implying that they could be part of a homogenous population.    

The spectrum of the accretion column emission is empirically described by a very hard power law (spectral index ${\lesssim}1.8$) with a high-energy exponential cutoff (${\sim}7-30$\,keV) \citep[e.g.,][and references therein]{2012MmSAI..83..230C}. Theoretical considerations indicate that the characteristic spectral shape is the result of bulk and thermal Comptonization of bremsstrahlung, blackbody, and cyclotron seed photons \citep{1981ApJ...251..288N,1985ApJ...299..138M,1991ApJ...367..575B,2007ApJ...654..435B}.  X-ray pulsars, are often modeled using simple phenomenological models that successfully describe the overall shape of the emission. Such empirical models work well on hard X-ray datasets ($>$10\,keV) with lower photon counts, but offer little insight to the physical processes that produce it. In the advent of NuSTAR which provides low background, high signal to noise X-ray spectra in the 3-80\,keV range and with the development of the analytic bulk and thermal Comptonization models by \cite[][henceforth BW]{2007ApJ...654..435B} the opportunity is presented to analyze XRP spectra with a more meticulous and physically motivated approach. 

The XRP \src that was recently observed during an outburst is an ideal target for this exercise. It is a galactic source that can yield spectra with large photon counts for moderate duration observations and as it has never before been observed by high resolution X-ray telescopes, we can provide the first insights into its  physical accretion flow parameters. The source was first discovered in the XMM slew survey \citep[source XMMSL1~J184555.4-003941][]{2008A&A...480..611S}, and its first major outburst was detected by MAXI on October 2019 \citep{2019ATel13189....1N}. Pulsations in its X-ray lightcurve were detected by \swift/XRT indicating a $\sim200$\,s spin period \citep{2019ATel13195....1K}. \swift/XRT also improved on the source localization, indicating 2MASS\,J18455462-0039341 as its optical counterpart \citep{2019ATel13218....1S,2019ATel13219....1K}, which was spectroscopically identified as a young Be-star \citep{2019ATel13211....1M,2019ATel13222....1M}, classifying \src as a Be X-ray binary. 

\src was observed by NuSTAR on Oct. 21 2019 for 23.5\,ks. A ${\sim}1$\,ks observation by \swift was also carried out on the same date. Here we report on the spectral and temporal characteristics of this newly discovered Galactic Be-XRB. We focus on the NuSTAR observation, as due to the high extinction towards the source \citep{2019ATel13222....1M} and the very short duration of the \swift observation no significant information can be collected below 3\,keV. One of the main motivations for this work is to exploit the high source count rate, thanks to its relatively small distance, which allows us to use more intricate and physically motivated models for the spectral analysis. We will use the bulk and thermal Comptonization model BW, to probe the values of fundamental physical parameters of the process of accretion onto a high-B neutron star. Namely, B-field strength, electron temperature, polar cap size and inferred mass accretion rate. 

\section{Data extraction \& analysis}
\label{sec:analysis}

We extract and analyze the spectra and light-curve of the NuSTAR observation of \src (see Table \ref{tab:log}). We model the source continuum with typical phenomenological models to confirm the characteristic spectral shape expected from radiation from the accretion column, and also the latest version of the BW accretion column emission model. We also inspect the source spectrum for the presence of emission and/or absorption-like features such as iron \ka emission and CRSF absorption lines.

\begin{table}
\caption{\nus observing log}
\label{tab:log}
\begin{threeparttable}
\begin{tabular*}{\columnwidth}{p{0.4\columnwidth}p{0.55\columnwidth}}
\hline
\hline\noalign{\smallskip} 
ObsID & 90501347002 \\
Target & Swift\_J1845d7m0037 \\
Start date & 2019-10-21 00:06:09\\
MJD start &  58777 \\
Exposure$^{(a)}$ & 23517 s \\
Time span$^{(b)}$ & 43864 s \\
\hline
\hline\noalign{\smallskip} 
\end{tabular*}
\tnote{(a)} Combined exposure of good time intervals. 
\tnote{(b)} Time span elapsed between first and last event detected.
\end{threeparttable}
\end{table}

\subsection{NuSTAR data extraction}

For the \nus data extraction, we used version 1.9.3 of the \nus data analysis system (\nus DAS) and the latest instrumental calibration files from CalDB v20191008. Date were cleaned and calibrated using the \texttt{NUPIPELINE} routine with default settings. reducing internal high-energy background was reduced and  passages through the South Atlantic Anomaly were screened (settings SAACALC$=$3, TENTACLE$=$NO and SAAMODE$=$OPTIMIZED). Phase-averaged source and background spectra were extracted using the \texttt{NUPRODUCTS} script, which also produces the instrumental responses for both focal plane modules, FPMA and B. We used a circular extraction region with an 80{\arcsec}  for both the source and the background spectra. The latter were extracted from a blank sky region in the same detector as the source and at adequate distance from it in order to avoid any contribution from the PSF wings. The default PSF, alignment, and vignetting corrections were used. 

\subsection{Temporal analysis}

\begin{figure}
 \includegraphics[width=\columnwidth]{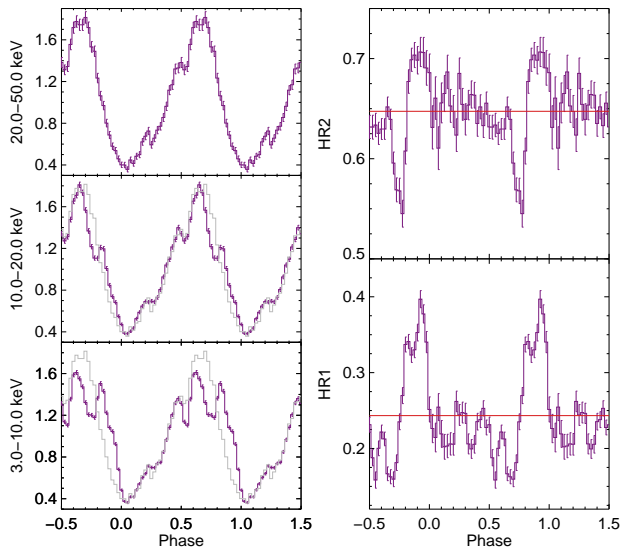}
 \caption{{\emph Left panels:} Pulse profiles of \src in 3 energy bands. In the 2 lower panels we have over-plotted (gray line) the pulse-profile of the 20-50 keV for visual comparison. {\emph Right panels:} We plot the phase resolved HR in the soft (HR1: 3.0--10.0 keV vs. 10.0-20.0 keV) and hard bands (HR2: 10.0--20.0 keV, 20.0--50.0 keV).}
 \label{fig:PP_HR}
\end{figure}

To search for pulsations in the \nus we used the combined events (both FPMA and B) obtained within the $\sim$24 ks exposure. We then searched for a periodic signal in the barycentric corrected event time of arrivals.
To derive an accurate pulse period measurement we performed an epoch folding test \citep{1983ApJ...272..256L} to events with energies between 3.0-60.0 keV (channel \# 47-1000). The test was implemented through \verb!python! by using \verb!stingray! and \verb!HENDRICS! \citep{Huppenkothen2019}.
A coherent periodic signal was found in the complete data set and in smaller segments of data. We used \verb!HENDRICS! {\tt HENphaseogram} command to calculate time of arrivals (ToAs) for individual pulses, and then fitted them with a Bayesian linear regression model \citep{2007ApJ...665.1489K}. We thus computed $P_{\rm NS}=207.38\pm0.03$~s.
The pulse profiles in 3 energy bands are presented in Fig. \ref{fig:PP_HR}. From the figure it is evident that the pulse profile becomes smoother at higher energies (i.e., 20.0-50.0 keV) where it approaches a single peaked structure, while as we move to lower energies two additional narrow peaks gradually appear around the main pulse peak.  

To track the spectral evolution with pulse phase we used the hardness ratio (HR), defined as $\rm{HR}=(\rm{R}_{i+1}-\rm{R}_{i})/(\rm{R}_{i+1}+\rm{R}_{i})$,   
where $\rm{R}_{i}$ is the count rate in a specific energy band. 
We split the phase folded events into three energy bands (i.e., 3.0--10.0 keV, 10.0--20.0 keV and 20.0--50.0 keV) and computed two HR indices using 40 phase bins. In Fig. \ref{fig:PP_HR} we plot HR as a function of the pulse profile. 

\begin{figure*}
 \resizebox{\hsize}{!}{
 \includegraphics[width=\columnwidth]{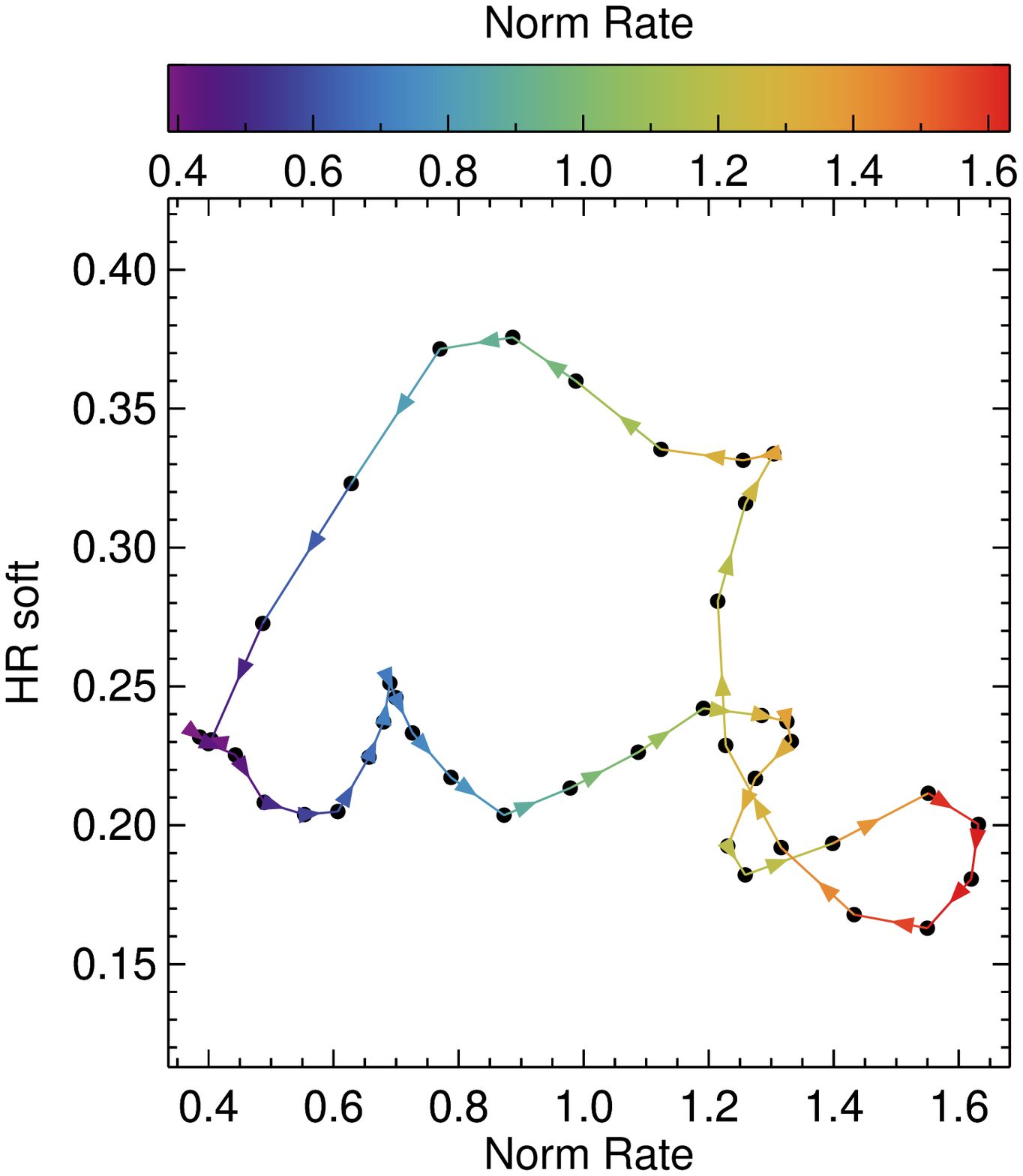}
 \includegraphics[width=\columnwidth]{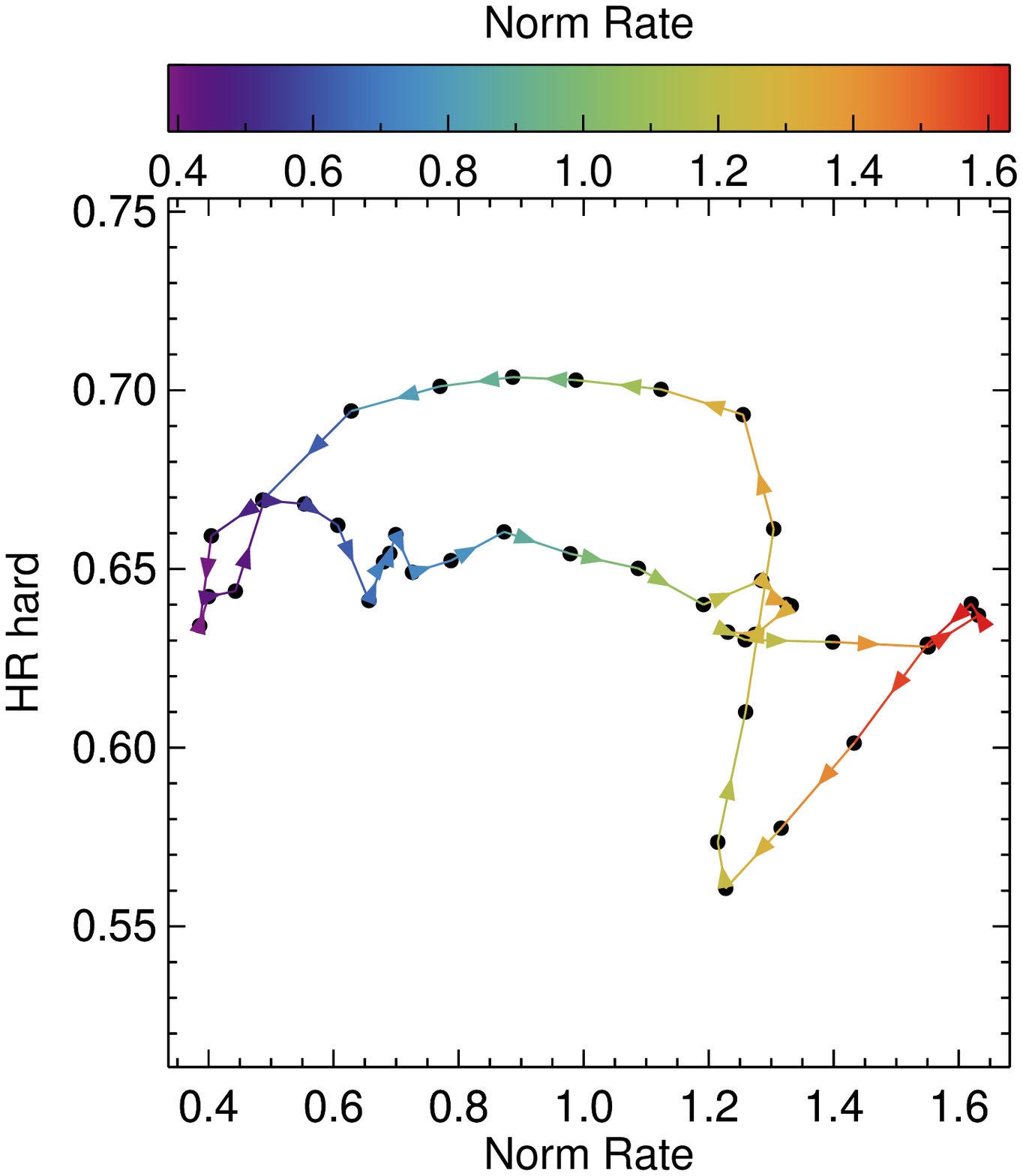}
 \includegraphics[width=\columnwidth]{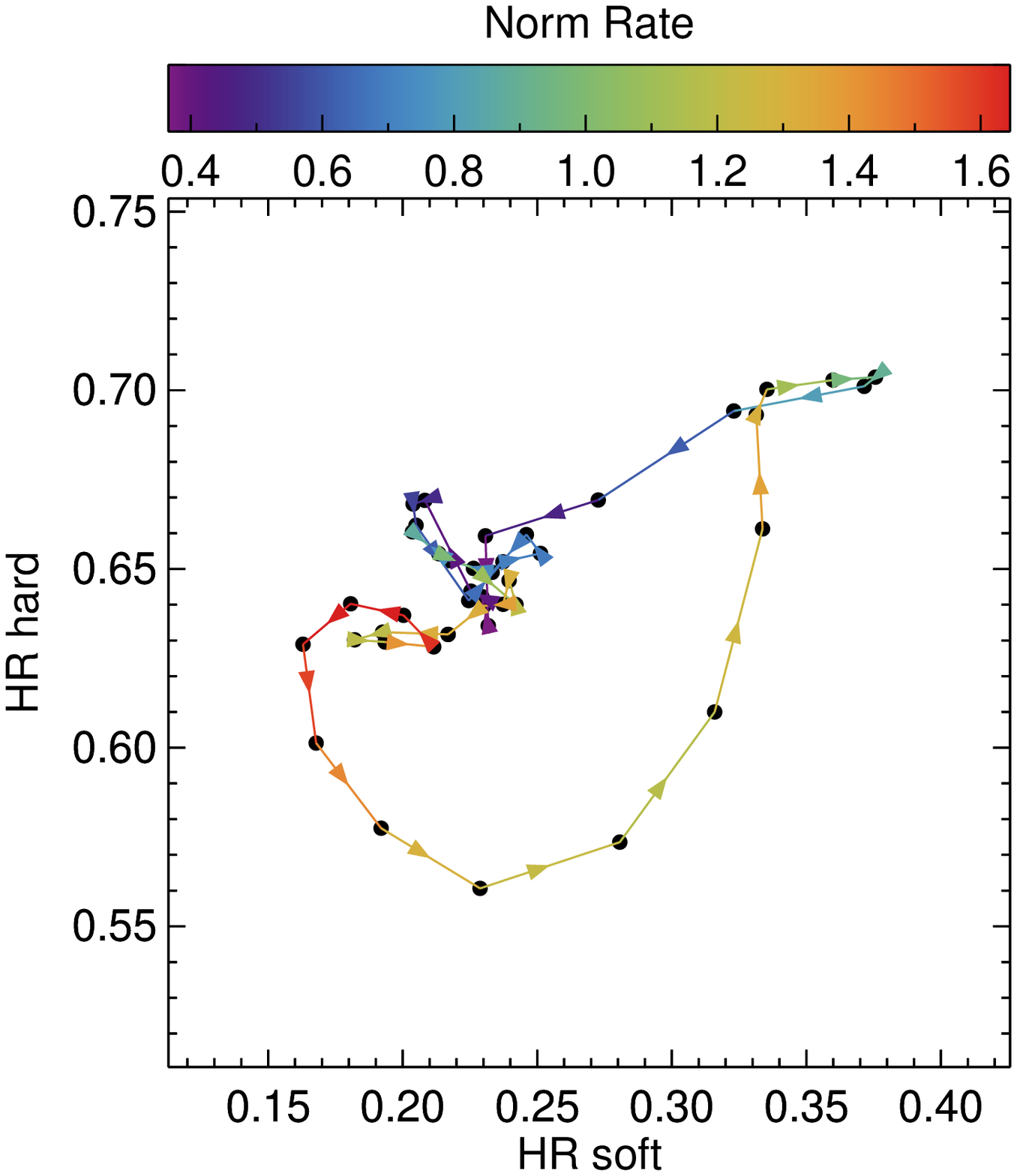}
 }
 \caption{Pulse profiles and HRs. Just to ilustrate HR evolution with Lx, Energy bands are the same as Fig. \ref{fig:PP_HR}.}
 \label{fig:PP_pulse}
\end{figure*}


\subsection{X-ray spectral analysis}

\begin{figure}
 \includegraphics[angle=-90, width=\columnwidth]{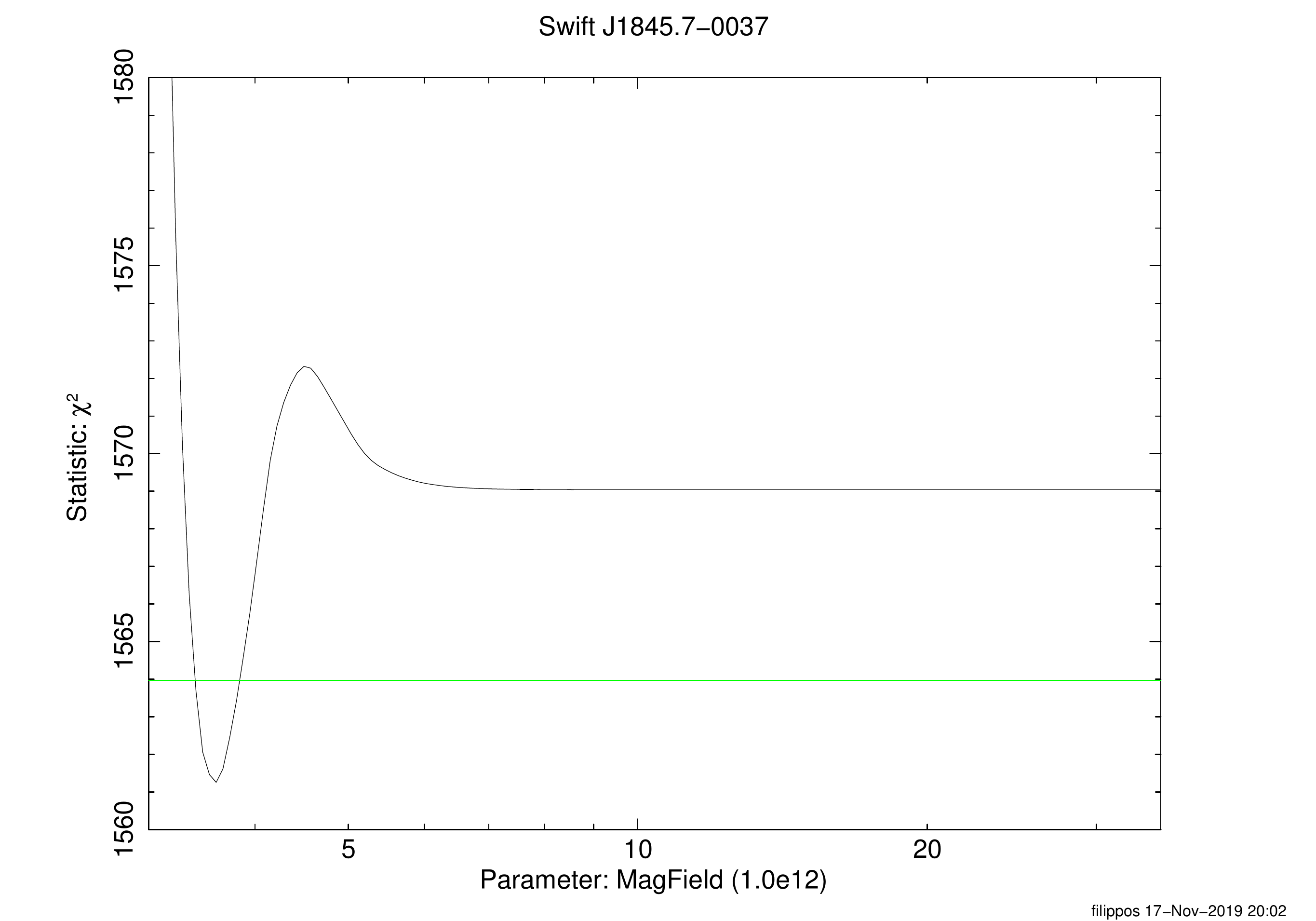}
 \caption{ Steppar for magnetic field strength. The parameter is constrained within the 20\,${\sigma}$ confidence range.}
 \label{fig:B_contour}
\end{figure}

\begin{figure}
 \includegraphics[angle=-90, width=\columnwidth]{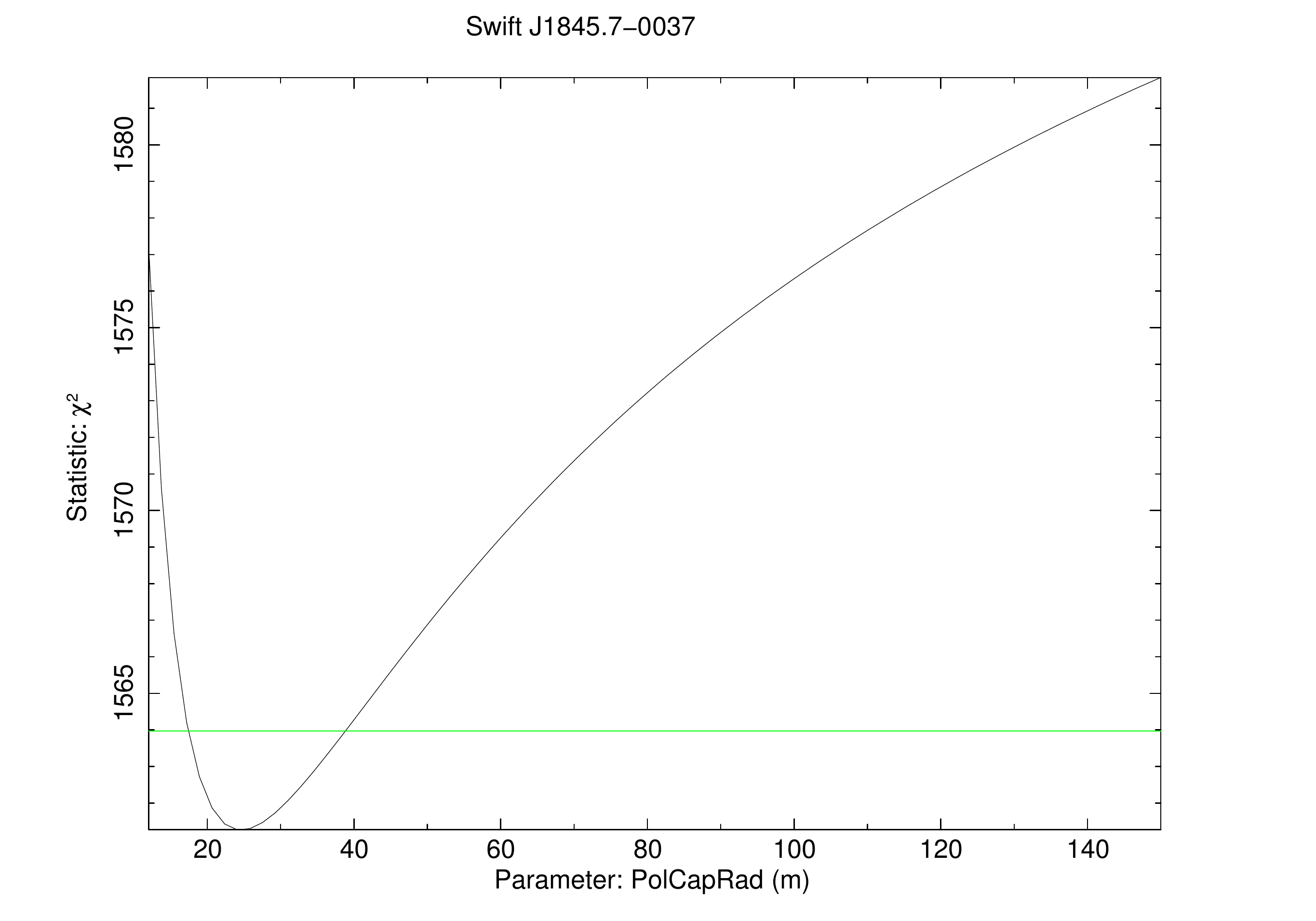}
 \caption{  Steppar for magnetic field strength. The parameter is constrained within the 20\,${\sigma}$ confidence range.}
 \label{fig:Ro_contour}
\end{figure}

\begin{figure}
 \includegraphics[angle=-90, width=\columnwidth]{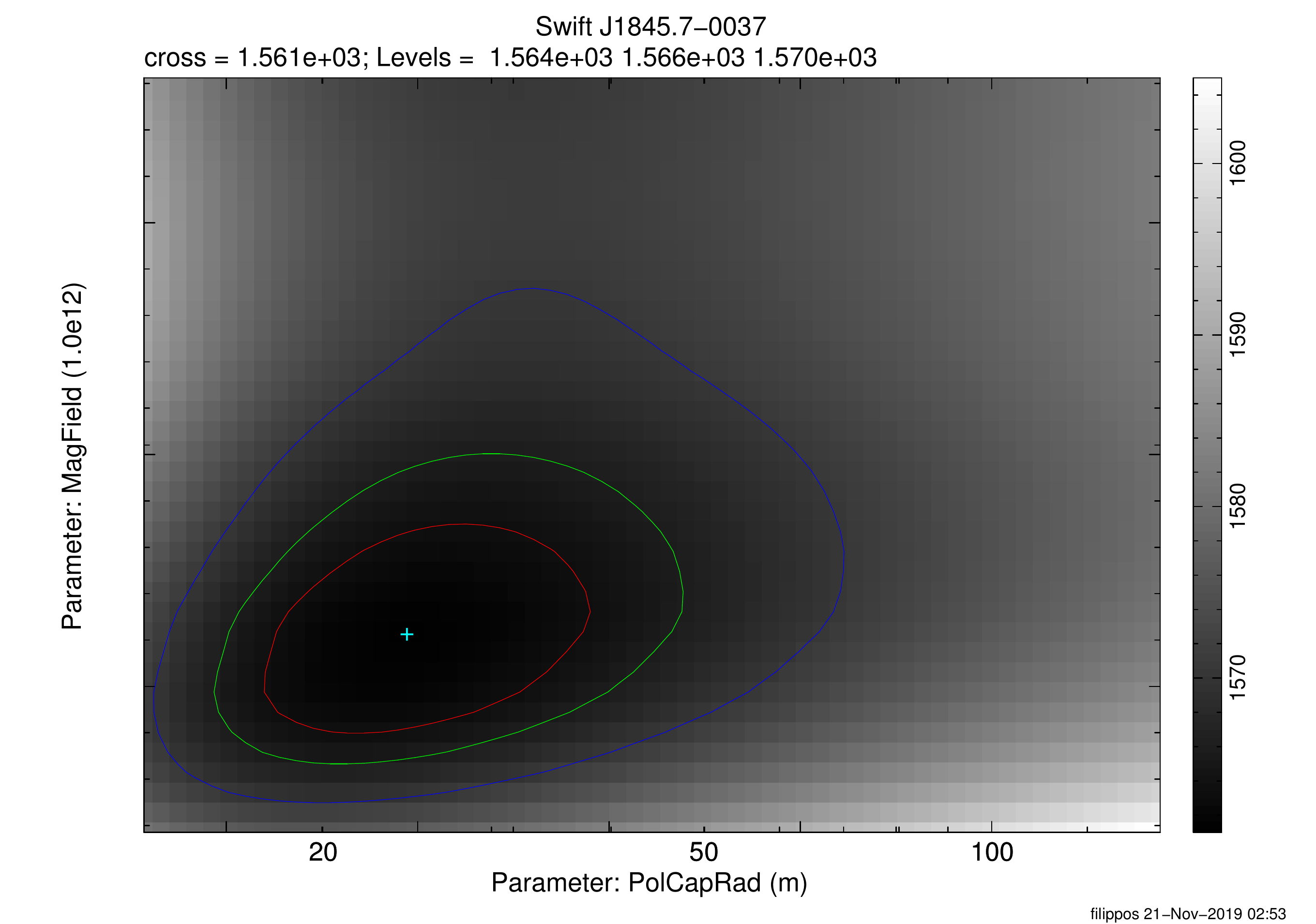}
 \caption{ 2D steppar for $B$ vs $R_o$}
 \label{fig:2D}
\end{figure}

\begin{figure}
 \includegraphics[angle=-90, width=\columnwidth]{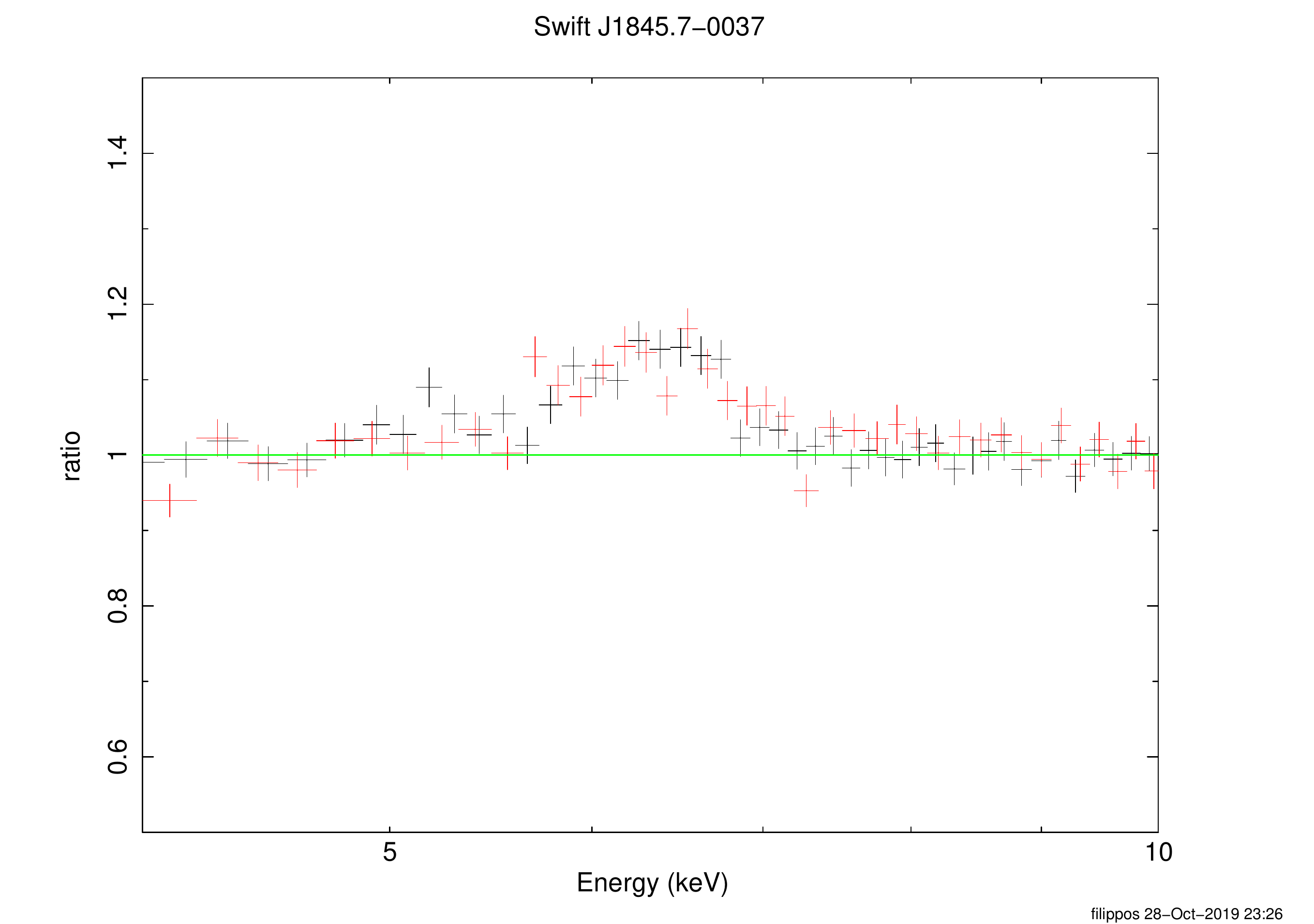}
 \caption{ Data-to-model ratio plots for the  Fe K${\alpha}$ emission line in the 4-10\,kev range. }
 \label{fig:Fe_ratio}
\end{figure}

\begin{figure}
 \includegraphics[angle=-90, width=\columnwidth]{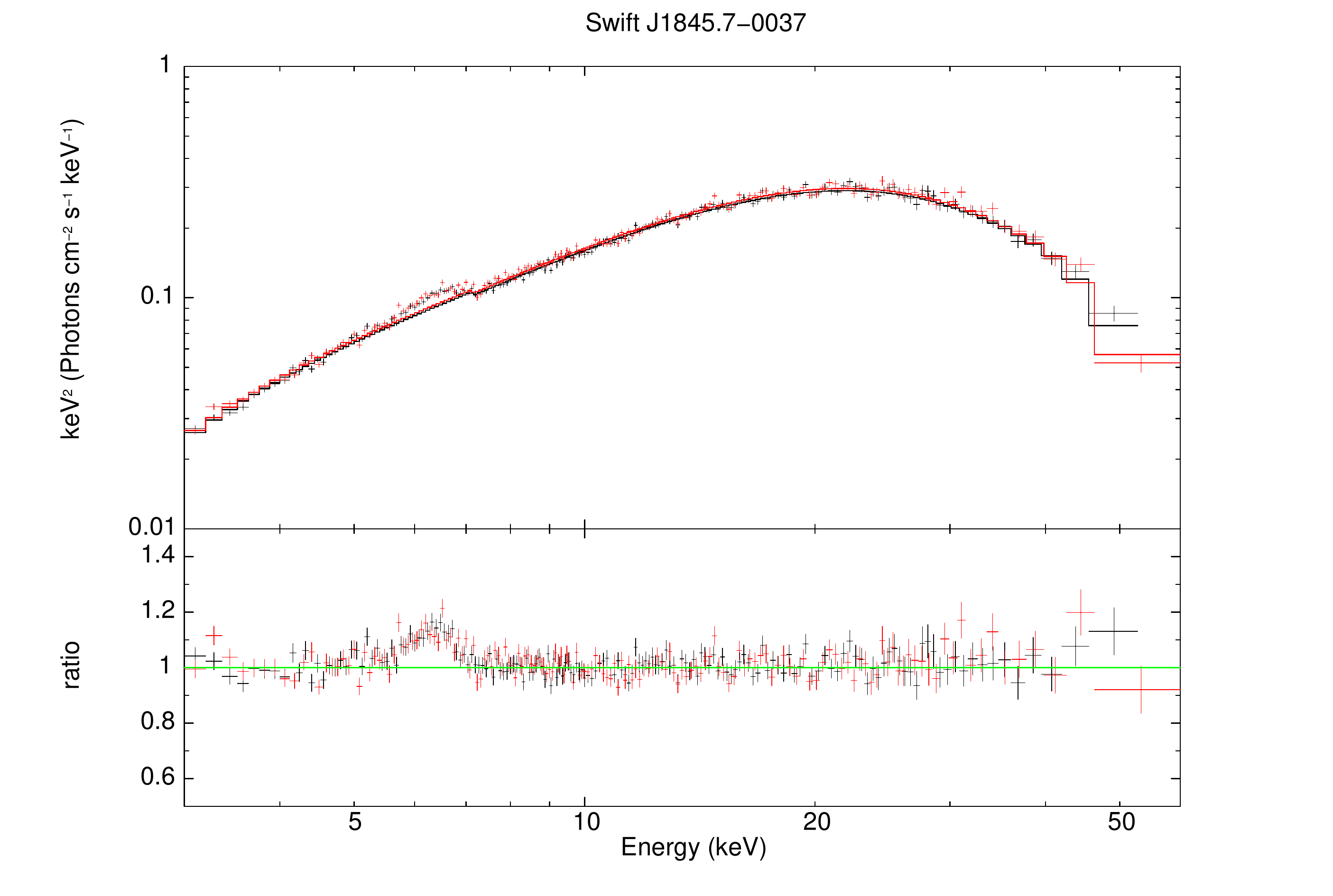}
 \caption{ Complete absorbed BW model without Gaussian emission line at ${\sim}$6.4. Unfolded spectrum just to see how it looks. }
 \label{fig:Fe_eeu}
\end{figure}

\begin{figure}
 \includegraphics[angle=-90, width=\columnwidth]{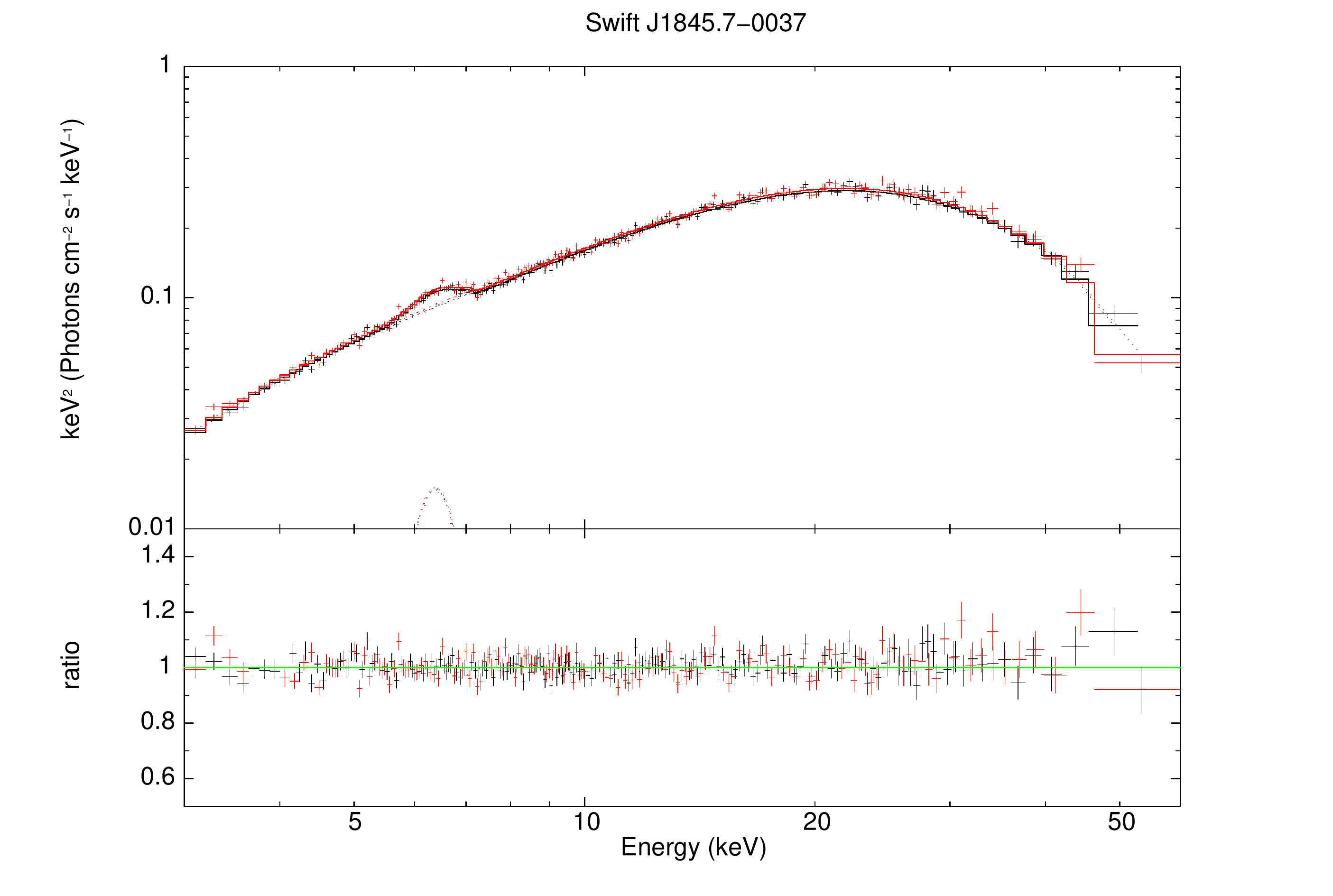}
 \caption{ Data-to-model ratio plot  for the absorbed BW model plus a Gaussian line for the Fe K${\alpha}$ emission line.}
 \label{fig:BW_eeu}
\end{figure}

\begin{figure}
 \includegraphics[angle=-90, width=\columnwidth]{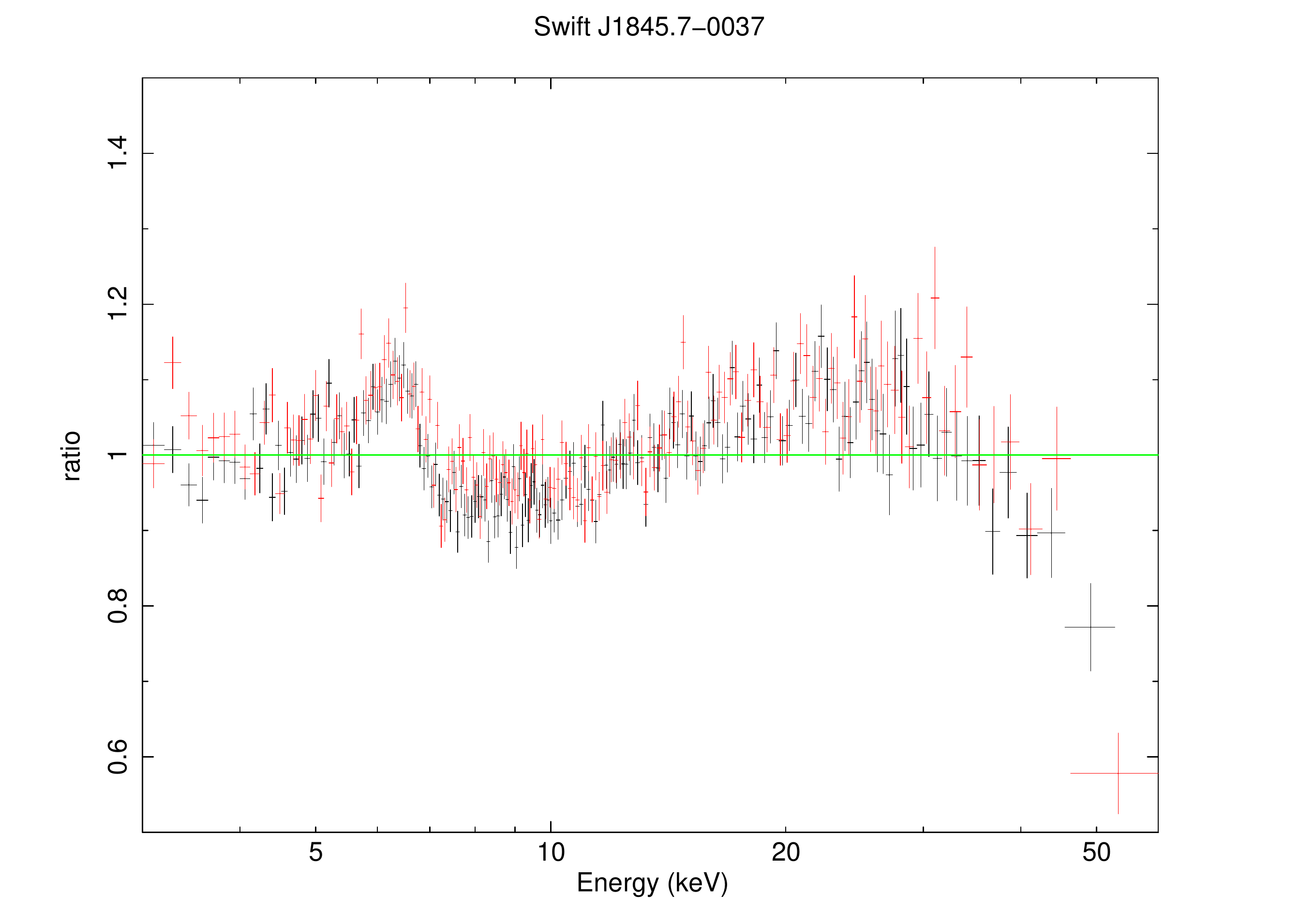}
 \caption{ Data-to-model ratio plot for a the absorbed \fdcut model.}
 \label{fig:therm_r}
\end{figure}

\begin{figure}
 \includegraphics[angle=-90, width=\columnwidth]{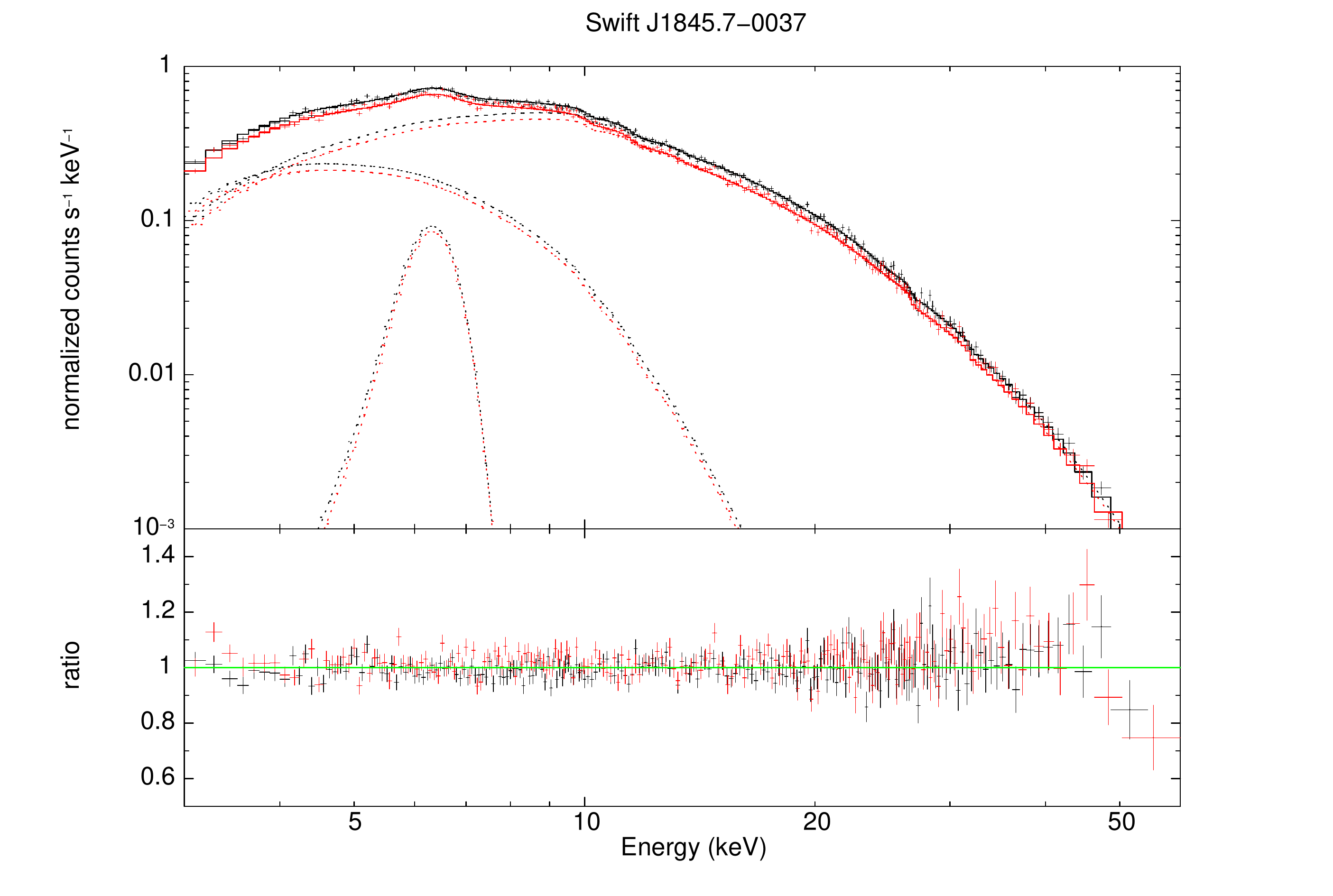}
 \caption{ Data-to-model ratio plot for a the empirical model.}
 \label{fig:empirical}
\end{figure}

\subsubsection{Empirical model}

We perform the X-ray spectral analysis using the {\tt Xspec} spectral fitting package, version 12.9.0 \citep{1996ASPC..101...17A}. To indicate basic spectral characteristics we employed standard models often used to empirically describe thermal and non-thermal emission from X-ray binaries of all types. More specifically, we employ a power-law with a high energy exponential cutoff to model non-thermal emission of the accretion column. We opted not to use  the   {\tt Xspec} model \texttt{cutoffpl} or a power-law multiplied with exponential cutoff model \texttt{highEcut}, as the former does not allow for cutoff steepness adjustment (i.e.~a "folding energy" parameter, $E_{\rm f}$) and the latter produces an abrupt break at the "cutoff energy", $E_{\rm c}$. Instead we used, what is known as the Fermi-Dirac cutoff model, usually denoted by \fdcut due to its apparent similarity to the Fermi-Dirac distribution. We note that this similarity is completely superficial and it is only an empirical expression with no physical connection to the F-D distribution. The use of the \fdcut expression, to describe high energy X-ray pulsar spectra, was first proposed by \cite{1986LNP...255..198T}, in order to improve upon the issues described above. It is given by,
\begin{equation}
F(E)=A\,E^{-\alpha}{\times}\frac{1}{+exp[(E-E_{\rm c})/E_{\rm f}]}     
\end{equation}
For any potential thermal emission we used the \texttt{bbodyrad} model. Emission or absorption lines were modelled with Gaussians. The interstellar absorption was modeled using the improved version of the \texttt{tbabs} code\footnote{http://pulsar.sternwarte.uni-erlangen.de/wilms/research/tbabs/} \citep{2000ApJ...542..914W}. The atomic cross-sections were adopted from \cite{1996ApJ...465..487V}. A sole cutoff power law does cannot adequately model the data, yielding a reduced ${\chi}^2$ value of 1.4 for 1522 dof, revealing considerable residual structure in the  data-to-model ratio plot (Figure \ref{fig:therm_r}). Narrow positive residuals -- consistent with iron the \ka emission line -- are also prominently detected in the data-to-ratio plot (Figures~\ref{fig:therm_r} and \ref{fig:Fe_ratio}). The addition of thermal component and a Gaussian emission line, significantly improve the fit quality, yielding a reduced ${\chi}^2$ value of 1.04 for 1515 dof, which we consider acceptable. Best-fit values for this model (henceforth referred to as the "empirical" model) are presented in Table \ref{tab:empirical}. We do not detect any absorption-like features consistent with Cyclotron Resonance Scattering (CRSF). While the empirical model produces a statistically acceptable fit of the data, there are still some residuals in the $>40$\,keV range (Figure \ref{fig:empirical}).

\subsubsection{Bulk \& thermal Comptonization model}

Based on our current understanding of mass accretion onto high-B neutron stars \citep[e.g.][ and references therein]{2012MmSAI..83..230C}, we expect that the spectral shape of the entire X-ray emission of \src in the NuSTAR band ((3-70\,keV) is the result of bulk and thermal Comptonization of Bremsstrahlung, Cyclotron and thermal seed photons inside the accretion column of the X-ray pulsar. To model this emission we used the latest version of the BW model combined with the \texttt{tbabs} model for the interstellar absorption. 
BW yielded a red.~${\chi}^2$ value of 1.03 for 1522 dof which is an improvement of ${{\delta}{\chi}}^2$ of 50 for the same number of free parameters. We note however that model  comparison should be done more rigorously using Bayesian model selection techniques. Such an endeavor is not within our scope. The best fit values for the BW model are presented in table \ref{tab:BW}. All parameters are well constrained and within expected values for X-ray pulsars. The models indicates a magnetic field strength of the order of $10^{12}$ and a polar cap radius of the order of 30\,km. The values are well constrained within a confidence range that exceeds 20\,${\sigma}$ (see Figures \ref{fig:B_contour}, \ref{fig:Ro_contour}). The 2D 1,2 and 3\,${\sigma}$ contour plots of $B$ vs polar cap size ($r_0$), further demonstrate the robustness of the spectral fit (Figure~\ref{fig:2D}).

\begin{table}
 \label{tab:empirical}
\caption{ Best-fit parameters of the empirical model. All errors are 1${\sigma}$.}
\begin{tabular}{lc}
  \hline  
    \hline
  Parameter              &  Value     \\
  
    \hline  
        \hline  
        \\
$N_{\rm{H}}$ (${\times}10^{22}$cm$^{-2}$) &$3.28_{-0.98}^{+0.82}$ \\
k${\rm T_{BB}}$ (keV)                 &$1.29{\pm}{0.20} $\\
{${\rm R_{BB}}^{a}$} (km)             &$0.43_{-0.01}^{+0.02}$\\
{${\rm \Gamma}$}                      &${0.59}{\pm}{0.03}$ \\
${\rm E_{c}}$ (keV)                 &$20.7_{-0.81}^{+0.77}$ \\
${\rm E_{f}}$ (keV)                &$8.94{\pm}{0.14}$ \\
${\rm norm_{PO}}$ (${\times}10^{-3}$) &6.46${\pm}{0.43}$\\
\\
$E_{\rm Fe}$ (keV)                    &$6.34{\pm}{0.03}$ \\
$\sigma_{\rm Fe}$ (keV)               &$0.38{\pm}{0.04}$ \\
${\rm norm_{\rm Fe}}$ (photons\,$10^{-3}$\,cm$^{-2}$\,s$^{-1}$)            &$0.37{\pm}{0.04}$ \\
\\
C$_{\rm FPM}$                         &$1.02_{-0.001}^{+0.002}$ \\
\\
red.~$\chi^2 / {\rm dof}$             &$1.04/1515$ \\
\\
$\rm Flux(3-60\,{\rm keV})$ (ergs ${\rm cm}^{-2}$ ${\rm s}^{-1}$) &$7.40{\pm}{0.01}{\times}10^{-10}$\\
$\rm L(3-60\,{\rm keV})$ (ergs\,${\rm s}^{-1}$)                   &$9.06{\pm}{0.10}{\times}10^{36}$\\
$\dot{M_{L}}^{a}$ ($10^{16}$g\,s$^{-1}$)       &$4.05{\pm}0.01$ \\
\\ 
  \hline  
  \hline  
\end{tabular}
{$^{a}$}{Inferred from luminosity, assuming and efficiency of 0.2} 
\end{table}

\begin{table}
 \label{tab:BW}
\caption{ Best-fit parameters of the absorbed BW model. All errors are 1\,${\sigma}$}
\begin{tabular}{lc}
  \hline  
    \hline
  Parameter              &  Value     \\
  
    \hline  
        \hline  
        \\
$N_{\rm{H}}$ (${\times}10^{22}$cm$^{-2}$) &$4.08{\pm}{0.06}$ \\
Distance (kpc)                        &10.0$^{a}$ \\
$\dot{M}$ ($10^{16}$g s$^{-1}$)       &$3.26_{-0.69}^{+0.61}$ \\
$kT_e$ (keV)                          &$5.59_{-0.17}^{+0.15}$\\
$r_0$ (m)                             &$26.5_{-7.36}^{+13.0}$\\
$B$ (Gauss)                           &${3.65_{-0.09}^{+0.15}}{\times}10^{12}$ \\
${\rm M_{NS}}$ (${\rm M_{\odot}}$)    &$1.4^{a}$  \\
${\rm R_{NS}}$ (km)                   &10$^{a}$ \\
$\sigma_{\perp}$ ($\sigma_{\rm{T}}$)  &1.0$^{a}$ \\
${\xi}$                               &$1.83_{-0.19}^{+0.12}$ \\
${\delta}$                            &$0.26{\pm}0.01$ \\
norm                                  &1.0$^{a}$  \\
\\
$E_{\rm Fe}$ (keV)                    &$6.33_{-0.02}^{+0.03}$ \\
$\sigma_{\rm Fe}$ (keV)               &$0.41_{-0.04}^{+0.05}$ \\
${\rm norm_{\rm Fe}}$ (photons\,$10^{-3}$\,cm$^{-2}$\,s$^{-1}$)            &$0.39{\pm}0.04$ \\
\\
C$_{\rm FPM}$                         &$1.02_{-0.001}^{+0.002}$ \\
\\
red.~$\chi^2 / {\rm dof}$             &$1.03/1515$ \\
\\
$\rm Flux(3-60\,{\rm keV})$ (ergs ${\rm cm}^{-2}$ ${\rm s}^{-1}$) &$7.39{\pm}0.01{\times}10^{-10}$\\
$\rm L(3-60\,{\rm keV})$ (ergs\,${\rm s}^{-1}$) &$9.07{\pm}0.003{\times}10^{36}$\\
$\dot{M_{L}}^{b}$ ($10^{16}$g\,s$^{-1}$)       &$4.05{\pm}0.01$ \\
\\ 
  \hline  
  \hline  
\end{tabular}
{$^{a}$}{Parameter frozen. \\}
{$^{b}$}{Inferred from luminosity, assuming an efficiency of 0.2} 
\end{table}

\section{Discussion}
\label{sec:discu}
The empirical modeling of the phased-averaged spectrum of \src reveals a spectral continuum shape consistent with typical spectra of XRPs. Namely, a hard power law with a high energy exponential cutoff and a broad and prominent iron \ka line. The addition of a thermal component was required in order to adequately fit the data. However, the presence of this component -- and more importantly its best-fit parameter values -- are difficult to reconcile with expected physical processes in accreting highly magnetized NSs. More specifically, the presence of hot thermal emission with kT in excess of 1\,keV from a region with a size of less than 2\,km is difficult to interpret in the context of a luminous X-ray pulsar ($L_{\rm X}>10^{36}$\,erg/s for a distance greater than 4\,kpc)  with an expected  large magnetosphere at which the accretion disk will be disrupted. For instance, if we assume a bipolar magnetic with magnetospheric radius of $R_{\rm m}{\sim}2{\times}10^7{\alpha}{\dot{M}_{15}}^{-2/7}{B_{9}}^{4/7}{M_{1.4}}^{-1/7}R_6^{12/7}$\,cm \citep{1977ApJ...217..578G,1979ApJ...232..259G,1979ApJ...234..296G,1992xbfb.work..487G} -- where ${\alpha}$  is a constant that depends on the geometry of the accretion flow, with ${\alpha}$=0.5 the commonly used value for disk accretion, ${\dot{M}_{15}}$ is the mass accretion rate in units of $10^{15}$\,g/s (estimated for the observed 3-60\,keV luminosity of ${\sim}$9.0${\times}10^{36}$\,erg/s and assuming an efficiency of 0.2), $B_{9}$ is the NS magnetic field in units of $10^{9}$ G, $M_{1.4}$ is the NS mass in units of 1.4 times the solar mass, and $R_6$ is the NS radius in units of $10^6$ cm -- we expect the $R_{\rm m}$ to range between 31 to 1612\,km for a B-field strength ranging between $10^9$ to $10^{12}$\,G, respectively (assuming a 1.4\,$M_{\odot}$ NS with a 10\,km radius). Therefore, the size of the emitting region does not correspond to our fit-inferred value. 
 
The bulk of the hard (${>}$4keV) X-ray emission of XRPs originates inside the accretion column. As the in-falling matter inside this funnel is stalled due to the emerging radiation it forms a thermal mound in a small area on the NS poles \citep{1973ApJ...179..585D}.
Optically thick thermal emission from the dense mound and cyclotron and Bremsstrahlung emission from its optically thin surface  must traverse the column of in-falling matter. Due to the extremely high temperatures the opacity along this path, is determined by the process of scattering. As a result the initial (seed) photons of the mound are scattered multiple times by in-falling electrons, undergoing both thermal and bulk Comptonization. The emerging radiation has a spectral shape that is generally described by a power law with a high energy exponential cutoff. This simple spectral form is subtly altered at lower energies (below 10keV) by the addition of the small fraction of seed photons that escape after few or no scatterings. With a sufficiently high photon count this excess emission can be detected and in the case of our empirical model it was represented by the hot (seemingly) thermal component. 

By employing the BW model the low energy residual structure was self-consistently modeled without the need for an additional component. The BW fit generates well constrained best-fit values of fundamental XRP parameters that are consistent with what is expected for a source of this type. The fit indicates that \src has a magnetic field with a strength of the order of $10^{12}$\,G and an accretion column with a radius of the order of 30\,m in the polar cap region. The mass accretion rate estimated by the BW model is approximately $10^{-9}\,{M{{\odot}}}$/yr in agreement with the luminosity inferred value (see Table ), assuming a 10\,kpc distance. The results of this exercise highlights the value of high quality hard X-ray observations of XRPs as probes of the physics of accretion column emission but also the capacity of the BW model to accurately describe XRP spectra and provide first indications of the principal parameters of these source. However, we stress that the cyclotron photons are injected as a delta function at the cyclotron energy so the comptonization of these photons is only roughly simulated within the BW model. The resulting value of the magnetic field strength is a suggestive estimation, in the absence of an observed CRSF.

Application of the BW  model to an increasing number of sources is crucial not only to provide insight to fundamental source parameters but also to probe the growing association between ULXs vs XRPs. Namely, while it evident that PULXs are accreting high-B pulsars the origin of their spectral shape -- which appears to be ubiquitous in the entire (non-pulsating) ULX population -- is hotly debated. Some authors have remarked that ULX spectral are XRP-like \citep[e.g.][]{2017ApJ...836..113P,2018ApJ...856..128W} others note that there are crucial distinctions between high-B NS persistently accreting above the Eddington limit (i.e.~PULXs) and typical XRPs which rarely and briefly exceed ${\sim}2{\times}10^{38}$\,erg/s \citep[e.g.][]{2017A&A...608A..47K,2017MNRAS.467.1202M,2019A&A...621A.118K}. Application of empirical models on low-count spectra do hint at superficial similarities between the spectra of numerous ULXs (pulsating or not) -- namely the hard emission can be modeled by a hard power-law with an exponential cutoff as in the works of \cite{2017ApJ...836..113P,2018ApJ...856..128W}. However, comparison of the best-fit values for the empirical model for \src -- a nominal Be-XRB in the sub-Eddington regime -- and all sources considered in the \citeauthor{2017ApJ...836..113P} and \citeauthor{2018ApJ...856..128W} reveals that the cutoff energy lies at ${\sim}20$\,keV, more than three times higher than the highest $E_{\rm c}$ value in the samples of both these works. Obviously, such elementary comparisons only serve as hints at discrepancies between the suggested common origin of the hard emission of PULX and X-ray pulsars.

A more rigorous approach is currently underway in which we attempt a systematic Bayesian comparison between all known X-ray pulsars in the ${\sim}10^{36}-10^{39}$\,erg/s range and (P)ULXs with available \nus data (Koliopanos et al.~in prep.). This project is already indicating a clear distinction between (P)ULX spectra and the accretion column emission of standard X-ray pulsars. An integral addition to the phenomenological analysis is the more nuanced spectral analysis of high quality spectra, using physical models, as illustrated in our application of the BW model on \src. Extending this analysis on a variety of sources at different luminosity regimes will probe the physics of radiative processes in accreting high-B NSs, but also the limitations of our theoretical framework, particularly in the super-Eddington regime.

Most studies that examine pulsar accretion structure use phase-resolved spectroscopy to examine how spectral parameters change as a function of pulsar spin phase. However, these analyses are model dependent and most spectral models used to describe NSs are phenomenological in nature (i.e.~cut-off power law) rather than physical. 
Such analysis was performed for \src by \cite{2019arXiv191101057D}, where the authors presented a phase dependence evolution of a typical {\tt fdcut} spectral component.   
The hardness ratios evolution with pulse phase can offer a model independent test of such phase-resolved spectroscopy results, that can be compared with physically motivated models.
\cite{2007ApJ...654..435B} developed a physically derived radiation driven radiative shock spectral model that successfully describes the spectra of bright X-ray pulsars and directly probes the accretion structure of NSs \citep{2016ApJ...831..194W}. 
Although the performance of phase resolved spectroscopy will be hindered by limited statistics (also model computes phase-averaged spectrum), comparison of model predictions with evolutionary paths of the HR (see Fig. \ref{fig:PP_pulse}) can potentially be used to predict changes in the observers angle with respect to the accretion column, as a function of pulse phase.


\section{Conclusions}
We have presented spectral and temporal analysis of \nus observation of the newly discovered Galactic Be-XRB \src. We have highlighted the pattern of  the pulse profile of the source in different energy ranges and explored the short-term temporal evolution of the source spectral hardness. We further utilized the high quality of the \nus observation to perform spectral analysis with both an empirical and  a physics-based model, employing the latest version of the BW model. This scheme allowed us to present the  basic phenomenological attributes of the source spectrum in the context of the established observational appearance of accreting high-B NSs, but also investigate the validity of -- often empirically claimed -- spectral components and probe the physical origin of the source emission. Employing the BW model allowed the assessment of intrinsic source parameters such as B-field strength and mass accretion rate, in the absence of more concrete evidence, such as CRSFs. This exercise, provides an additional exhibition of the capacity of the bulk and thermal Comptonization model as an effective tool to further identify the radiative signatures of physical processes within the accretion column of X-ray pulsars.

\label{sec:conclu}

%
\bibliographystyle{aa}
\bibliography{general}

%

\end{document}